\documentclass[final,12pt]{elsarticle}

\usepackage{amssymb}
\usepackage{amsmath}

\usepackage{xcolor}

\journal{Journal of High Energy Astrophysics}

\def\prl{Phys. Rev. Lett. }
\def\prd{Phys. Rev. D }
\def\aap{A\&A }
\def\apj{ApJ }
\def\apjl{ApJ }

\def\mnras{MNRAS }

\begin{document}

\begin{frontmatter}



\title{Short GRB 090510: a magnetized neutron star binary merger leading to a black hole}

\author[icranetpescara,icra,icranetferrara,unife,inafrome]{J. A. Rueda} 
\ead{jorge.rueda@icra.it}

\author[icranetpescara,icra,inaf]{R. Ruffini} 
\ead{ruffini@icra.it}

\author[icranetpescara,icra,inafteramo]{Yu Wang}
\ead{yu.wang@icranet.org}


\affiliation[icranetpescara]{organization={ICRANet},
            addressline={Piazza della Repubblica 10}, 
            city={Pescara},
            postcode={I-65122}, 
            country={Italy}}

\affiliation[icra]{organization={ICRA},
            addressline={Dipartimento di Fisica, Sapienza Università di Roma}, 
            city={Rome},
            postcode={I-00185}, 
            country={Italy}}

\affiliation[icranetferrara]{organization={ICRANet-Ferrara, Dipartimento di Fisica e Scienze della Terra, Università degli Studi di Ferrara},
            addressline={Via Saragat 1}, 
            city={Ferrara},
            postcode={I-44122}, 
            country={Italy}}

\affiliation[unife]{organization={Dipartimento di Fisica e Scienze della Terra, Università degli Studi di Ferrara},
            addressline={Via Saragat 1}, 
            city={Ferrara},
            postcode={I-44122}, 
            country={Italy}}

\affiliation[inafrome]{organization={INAF, Istituto de Astrofisica e Planetologia Spaziali},
            addressline={Via Fosso del Cavaliere 100}, 
            city={Rome},
            postcode={I-00136}, 
            country={Italy}}

\affiliation[inaf]{organization={INAF},
            addressline={Viale del Parco Mellini 84}, 
            city={Rome},
            postcode={I-00136}, 
            country={Italy}}

\affiliation[inafteramo]{organization={INAF,                        Osservatorio Astronomico                                d'Abruzzo},
            addressline={ Via M. Maggini snc}, 
            city={Teramo},
            postcode={I-64100}, 
            country={Italy}}
     
\begin{abstract}
We model the short gamma-ray bursts (GRB) 090510 as the product of a magnetized neutron star (NS) binary merger. Accounting for the NS critical mass constraint given by the mass of PSR J0952--0607, we infer that GRB 090510 was a highly-magnetized NS-NS merger that left as remnant a Kerr black hole (BH) of $2.4 M_\odot$ with a low-mass accretion disk. The gamma-ray precursor is powered by the magnetic energy released during the merger of the NSs. The prompt emission originates at the transparency of an ultra-relativistic $e^+e^-$ pair-plasma produced by the overcritical electric field induced by the rotating strong magnetic field around the merged object before it reaches the critical mass, the GeV emission by the extractable energy of the newborn BH, and the X-ray afterglow by accretion onto it. We derive the masses of the merging NSs, their magnetic fields, the BH mass, spin, and irreducible mass, the strength of the magnetic field, the disk mass, and obtain an estimate of the gravitational-wave emission during the merger phase preceding the prompt short GRB emission. The inferred parameters agree with up-to-date numerical relativity simulations, confirming that strong magnetic fields above $10^{14}$ G develop in NS-NS mergers and that mergers leading to a central BH remnant have low-mass disks of $\sim 10^{-2} M_\odot$. We also advance the possibility that quasi-period oscillations of tens of Hz of frequency due to Lense-Thirring precession of the matter surrounding the merged object before BH formation can explain the successive spikes following the prompt emission peak.
\end{abstract}





\end{frontmatter}
%

\section{Introduction} \label{sec:1}

The absolute upper limit to the mass of a non-rotating neutron star (NS) of $3.2 M_\odot$ has been established on the grounds of the existence of a fiducial nuclear density and theoretical constraints \cite{1974PhRvL..32..324R}. Recently, the lower limit of $2.35 M_\odot$ to the critical mass of a non-rotating NS has been established by the observation of the highest NS mass measured, that of PSR J0952--0607 \cite{2022ApJ...934L..17R}. This limit establishes the lower limit of the black hole (BH) in the binary-driven hypernova (BdHN) family of long gamma-ray bursts (GRBs) characterized by seven different emission Episodes (\cite{2023ApJ...955...93A} and references therein) and, as shown here, also constrains short GRBs. 

Since the early GRB literature, the nature of short GRBs has been amply recognized in merging NS-NS or NS-BH binaries (see, e.g., \cite{1986ApJ...308L..47G, 1986ApJ...308L..43P, 1989Natur.340..126E, 1991ApJ...379L..17N}). However, understanding their emission mechanisms and determining whether a BH forms remains challenging. To make one step forward in the knowledge of short GRBs, we here revisit our early interpretation of short GRB 090510 \cite{2016ApJ...831..178R}, based on theoretical and observational progress gained in the intervening 10 years, and fulfilling the most recent NS critical mass constraint \cite{2022ApJ...934L..17R}.

Summarizing, the picture that emerges {in the present analysis} for GRB 0905010 is that it was produced in a magnetized NS-NS merger. The merged central remnant surpassed the NS critical mass, so it promptly collapsed, forming a Kerr BH \cite{1963PhRvL..11..237K}. During the merger phase, the interaction of the strong fields releases the magnetic energy powering the GRB \textit{precursor}. Moreover, the high rotation of the surrounding magnetic field also induces a strong, overcritical electric field that creates an $e^+e^-$ plasma. The latter self-accelerates to an ultrarelativistic speed, leading to a succession of transparency events that radiate MeV photons \cite{1999A&A...350..334R,2000A&A...359..855R}, {which lead to } multiple blackbodies observed {in what we have called ultrarelativistic prompt emission (UPE) phase (see, e.g., \cite{2021PhRvD.104f3043M,2022EPJC...82..778R})}. While we confirm the data analysis that unveiled the successive blackbody emissions in the UPE of GRB 090510 \cite{2016ApJ...831..178R}, our early interpretation of it based on the $e^+e^-$ plasma produced by a charged BH \cite{1975PhRvL..35..463D,1998A&A...338L..87P} is here superseded by a change of paradigm, guided by the UPE analysis of long GRBs in the BdHN scenario \cite{2021PhRvD.104f3043M,2022EPJC...82..778R}, in which the $e^+e^-$ plasma is produced by a strong electric field induced by rapidly rotating magnetic field lines. We also propose and investigate the possibility that Lense-Thirring precession of the material around the merged object around the equatorial plane produce QPOs that lead to sequential spikes of tens of Hz of frequency, similar to those observed in the GRB 090510 light curve after the UPE, and which is left for further observational inquiry, confirmation, and exploration in additional sources. After the UPE, the residual electric field accelerates charged electrons, whose radiation leads to the observed GeV emission detected by the Fermi-LAT. The extractable energy of the BH is the energy reservoir of the GeV emission. The angular momentum in the merger is enough to form a small-mass accretion disk onto the BH, leading to the X-ray afterglow.  

This approach allows us to consistently infer the values of the masses of two merged NSs, the mass and spin of the newborn BH, and the strength of the magnetic field surrounding it. The relatively low X-ray emission originating from the accretion of the ejected mass in the merger process is comfortably explained by a negligible mass asymmetry of the two merged NSs. As a natural result of this analysis, we obtain the expected gravitational wave (GW) emission during the inspiral and merger, as well as the mass of the disk formed around the BH from the unbound material with non-zero angular momentum expelled in the merger process. 

\section{Observations}\label{sec:2}

GRB 090510 is a bright, short-hard gamma-ray burst detected by multiple instruments, including Fermi’s Gamma-ray Burst Monitor (GBM) and Large Area Telescope (LAT) \cite{2010ApJ...716.1178A}, the AGILE satellite \cite{2010ApJ...708L..84G}, and Swift’s Burst Alert Telescope (BAT) \cite{2009GCN..9331....1H}. Its T90 duration is approximately $0.3$ s, placing it within the short GRB class \cite{2010ApJ...716.1178A}. The redshift, determined from the host galaxy, is $z = 0.903$ \cite{2009GCN..9353....1R}.

The prompt emission light curve shows two distinct phases \cite{2013ApJ...772...62M}. The first phase, the \textit{precursor}, from rest-frame time T0 - $0.06$ s to T0 + $0.06$ s, is intense in the $0.3$--$10$ MeV range, with a spectral peak near $3$ MeV \cite{2010ApJ...708L..84G}. No significant gamma-ray emission above $30$ MeV is detected during this time interval. {We analyze the precursor emission in Section \ref{sec:3}.}

The second phase, from $0.25$ s to about $0.5$ s, is marked by the onset of gamma-rays above $30$ MeV, detected by LAT \cite{2010A&A...510L...7G}. Its spectrum is best fitted by the sum of a Band function and a power-law component extending from sub-MeV to GeV \cite{2010ApJ...716.1178A, 2010ApJ...708L..84G}. {This second phase of the prompt includes the UPE phase associated with the pair plasma transparency events, characterized by thermal emission, which we analyze in Section \ref{sec:4}.}

\begin{figure}
    \centering
    \includegraphics[width=\hsize,clip]{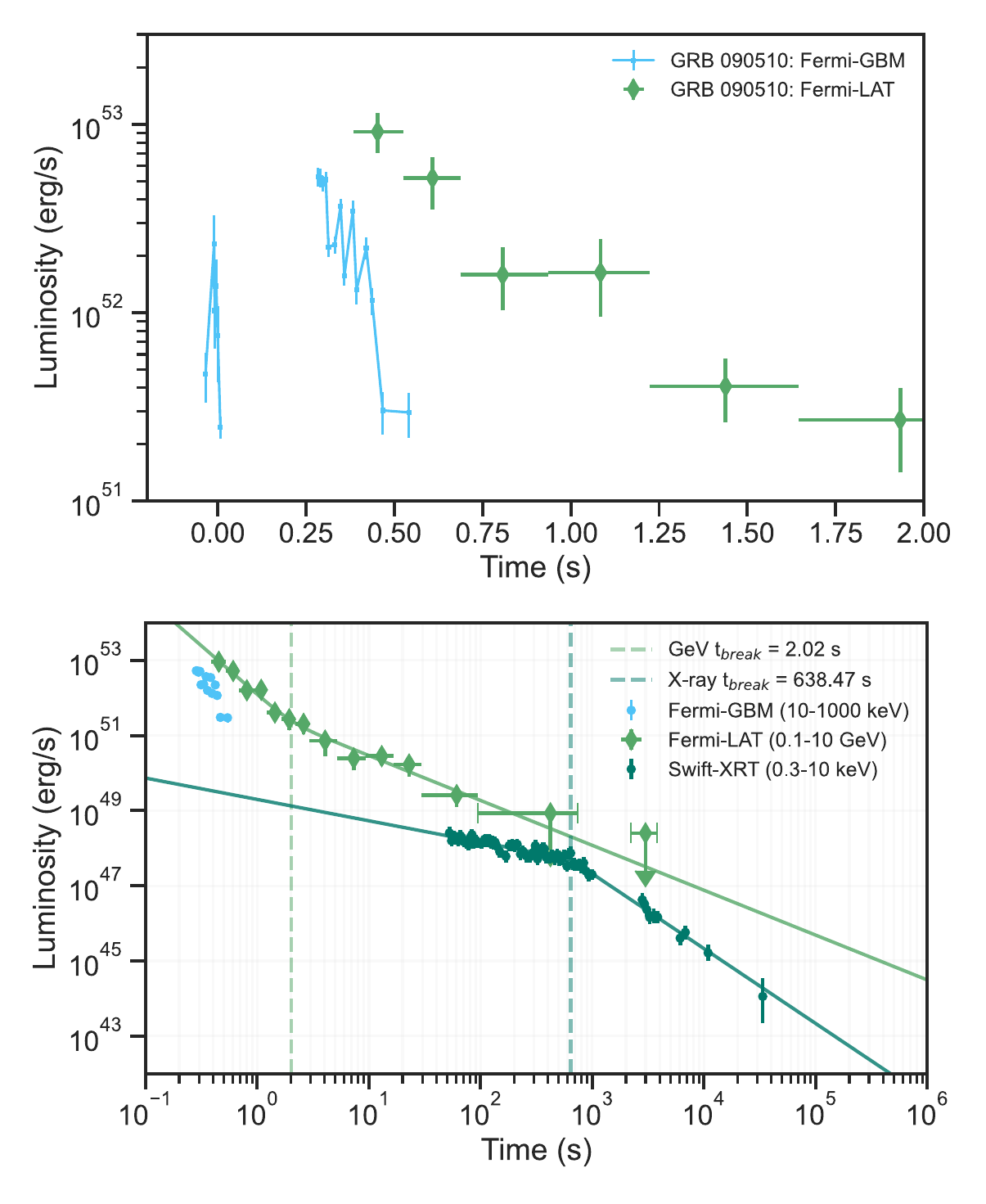}
    \caption{Rest-frame luminosity light curves of GRB 090510 observed by Fermi/GBM (blue), Fermi/LAT (green), and Swift/XRT (deep green). The upper panel shows the prompt emission within the first 2 s on a linear scale, where two distinct prompt phases are visible. The lower panel presents the full temporal evolution on a logarithmic scale, together with the smoothed broken power-law fitting.}
    \label{fig:lc}
\end{figure}

Significant GeV emission persists for up to $200$ s, and the highest energy photon ever detected from a short GRB, at $30.5$ GeV, arrived $0.829$ s after the GBM trigger \cite{2010ApJ...716.1178A}. We follow standard procedures to produce luminosity light curves for Fermi/GBM \cite{2012AAS...21914909G}, Fermi/LAT\cite{2013ApJS..209...11A}, and Swift/XRT \cite{2007A&A...469..379E}, shown in Fig \ref{fig:lc}. These processes and results are consistent with the analysis presented in our previous papers \cite{2013ApJ...772...62M, 2018JCAP...10..006R}. {We analyze the GeV emission and its modeling through the BH extractable energy in Section \ref{sec:5}.}

{We now} briefly summarize the data processing procedure:

\begin{itemize}
    \item GBM Data Analysis: We analyze GRB 090510 using Time-Tagged Event (TTE) data from the NaI 6 and BGO 1 detectors, selected for their optimal viewing angles to the burst. The energy ranges are $8$--$900$ keV for NaI 6 and $300$ keV--$40$ MeV for BGO 1. We perform time-resolved spectral fitting in XSPEC using the Band function. The luminosity is calculated as $L = 4\pi k D_L^2 F_{\rm obs} $, where $F_{\rm obs}$ is the observed flux, $D_L$ is the luminosity distance, and $k$ is the $k$-correction for the redshift and spectral shape, the Planck 2018 cosmological parameters \citep{2020A&A...641A...6P} are adopted.
    \item LAT Data Analysis: We process LAT data using the P8R3\_SOURCE event class and P8R2\_SOURCE\_V6 instrument response functions, in the $100$ MeV--$100$ GeV range. Data selection uses gtselect with standard cuts: DATA\_QUAL=1, LAT\_CONFIG=1, and zenith angle $< 90$ deg. We perform unbinned maximum likelihood analysis with gtlike, modeling a 10 deg region centered on the burst and including nearby catalog sources, as well as Galactic and isotropic diffuse backgrounds. The source spectrum is modeled as a power-law. We generate light curves with adaptive binning, requiring a minimum test statistic (TS) $>20$ per bin. Luminosity conversion uses the same formula as in the GBM analysis.
    \item XRT Data Analysis: XRT observations start about 100 s ($52.5$ s rest-frame time) after the BAT trigger, initially in Windowed Timing mode, then transitioning to Photon Counting mode. We use the automated XRT pipeline products from the UK Swift Science Data Centre, which apply corrections for bad columns, hot pixels, and pile-up. The photon index for each time bin is estimated from hardness ratio measurements, and a $k$-correction is computed accordingly. The observed $0.3$--$10$ keV flux is then converted to the time-resolved luminosity light curve using the same formula as above.
\end{itemize}

\section{The precursor}\label{sec:3}

Up-to-date numerical relativity simulations show that NS-NS mergers can amplify the NS merging fields to very high values that can reach even $10^{17}$ G
\cite{
2006Sci...312..719P,2025arXiv250410604A,2025PhRvD.111d4038B}. These extreme magnetic fields store sufficient energy to power a transient emission like the one observed in the precursor of GRB 090510, which has an isotropic-equivalent value of $E_{\rm prec} = 2.28\times 10^{51}$ erg \cite{2016ApJ...831..178R}, as shown in Fig.~\ref{fig:lc}, the first spike appears around $t = 0$. As shown in \ref{app:A}, the electromagnetic energy of the system is well approximated by the energy of the magnetic dipole (see Eq. \ref{eq:Wmagf}, and discussion therein)
\begin{equation}\label{eq:Umag}
    W \approx W_{\rm mag} = \frac{1}{4} B_p^2 R^3 {= 4.32\times 10^{49} \left( \frac{B_p}{10^{16}\,\rm G} \right)^2 \left( \frac{R}{12\,\rm km} \right)^3\,\,\text{erg}},
\end{equation}
where $B_p$ is the magnetic field strength on the stellar pole, i.e., for $r=R$ and $\theta = 0$. By equating the magnetic energy (\ref{eq:Umag}) to the precursor isotropic energy, and adopting a fiducial NS radius of $12$ km, consistent with most recent constraints from NICER data on the radius of a canonical NS (see, e.g., \cite{2025PhRvD.111c4005B}, and references therein), we obtain $B_p = 7.3\times 10^{16}$ G. This value can be adopted as an upper limit of $B_p$ at the beginning of the merger, since the emission can have some beaming. For instance, assuming $W_{\rm mag} = E_{\rm prec} (1-\cos\theta_{\rm prec})$, the inferred magnetic field strength is $B_p = 2 R^{-3/2}\sqrt{E_{\rm prec} (1-\cos\theta_{\rm prec})} = 8.50\times 10^{15}$--$2.66\times 10^{16}$ G for beaming angles in the range $\theta_{\rm prec} = 10^\circ$--$30^\circ$.

An estimate of the luminosity approaching the merger can be estimated via the magnetic flux across the region of open magnetic field lines, $\Phi_{\rm open} = B_p S_{\rm open}$, as follows \cite{2005A&A...442..579C,2006ApJ...643.1139C,2006ApJ...648L..51S}
\begin{equation}\label{eq:Lprec}
    L_{\rm prec} \sim L_{\rm mag} = \frac{\Omega^2}{6 \pi^2 c} \Phi_{\rm open} = \frac{\Omega^2}{6 \pi^2}B_p S_{\rm open} = \frac{1}{6} B_p^2 R^2 c \left( \frac{\Omega R}{c} \right)^4,
\end{equation}
where $S_{\rm open}$ is the area of the region of open magnetic field lines. We have used that, for the present rotating dipole, the radius of the open field lines region at the stellar pole is $l_{\rm open} \approx R \theta_{\rm open} \approx R (\Omega R/c)^{1/2}$, so its area is given by $S_{\rm open} \approx \pi l_{\rm open}^2 = \pi R^2 (\Omega R/c)$. Notice that Eq. (\ref{eq:Lprec}) has the same functional dependence of the luminosity by an orthogonal rotating dipole in vacuum. 

At the merger, $r = r_{\rm mgr} \approx 2 R$, $\Omega = \Omega_{\rm mgr} \approx (G M/r_{\rm mgr}^3)^{1/2}$, and Eq. (\ref{eq:Lprec}) gives the radius-independent value
\begin{equation}\label{eq:Lprec2}
    L_{\rm prec, mgr} = \frac{G^2}{384 c^3} (B_p M)^2 \approx 9.78\times 10^{50} \left(\frac{B_p}{10^{16}\,{\rm G}} \frac{M}{2.4 M_\odot}\right)^2\,\,\text{erg s}^{-1},
\end{equation}
Notice that this luminosity is consistent with the observed precursor luminosity, which is in the range of a few $\sim 10^{51}$--$10^{52}$ erg s$^{-1}$ (see Fig. \ref{fig:lc}), with a mild beaming angle similar to aforementioned ones. {The associated timescale of the energy release is 
\begin{equation}
    \tau_{\rm prec} = \frac{W_{\rm mag}}{L_{\rm prec}} = \frac{96 \,c^3 R^3}{G^2 M^2} = 44.17 \left( \frac{2.4\,M_\odot}{M} \right)^2 \left(\frac{R}{12\,\rm km} \right)^3\,\,\text{ms},
\end{equation}
which agrees with the timescale of the precursor, as shown by Fig. \ref{fig:lc}. Notice the above timescale is independent of the magnetic field strength, it depends only on the mass and the size of the system.
}

It is worth noting that the above estimates also agree with recent numerical simulations of the possible electromagnetic precursor from the magnetized NS-NS mergers. Additionally, it is also estimated that the radiation is likely to be in the MeV energy regime (see, e.g., \cite{2025arXiv250319884S}). 

{To get further insight into the precursor emission mechanism, e.g., a possible similarity with the UPE phase process, we checked the possible presence of a thermal emission in the precursor. Hence, we attempted to fit the precursor data from the Fermi-GBM n6 and n8 detectors, which are the triggered detectors, together with the corresponding b1 detector for MeV photons. Unfortunately, the available data are insufficient to determine whether a thermal component is present. The precursor has a photon flux of approximately $3$ photons s$^{-1}$ cm$^{-2}$ and an energy flux of about  $10^{-6}$ erg s$^{-1}$ cm$^{-2}$, which are relatively low. We first fitted the precursor with a power-law, cutoff power-law, and Band function. Then, for each of these models, we added a thermal component and repeated the fits. The inclusion of a thermal component reduced the C-statistic by only $1$--$3$, while also decreasing the degrees of freedom by two. Statistically, this reduction is not significant enough to confirm the presence of a thermal component.}

\section{The UPE phase}\label{sec:4}

\begin{figure}[hbtp]
    \centering
    \includegraphics[width=0.6\hsize,clip]{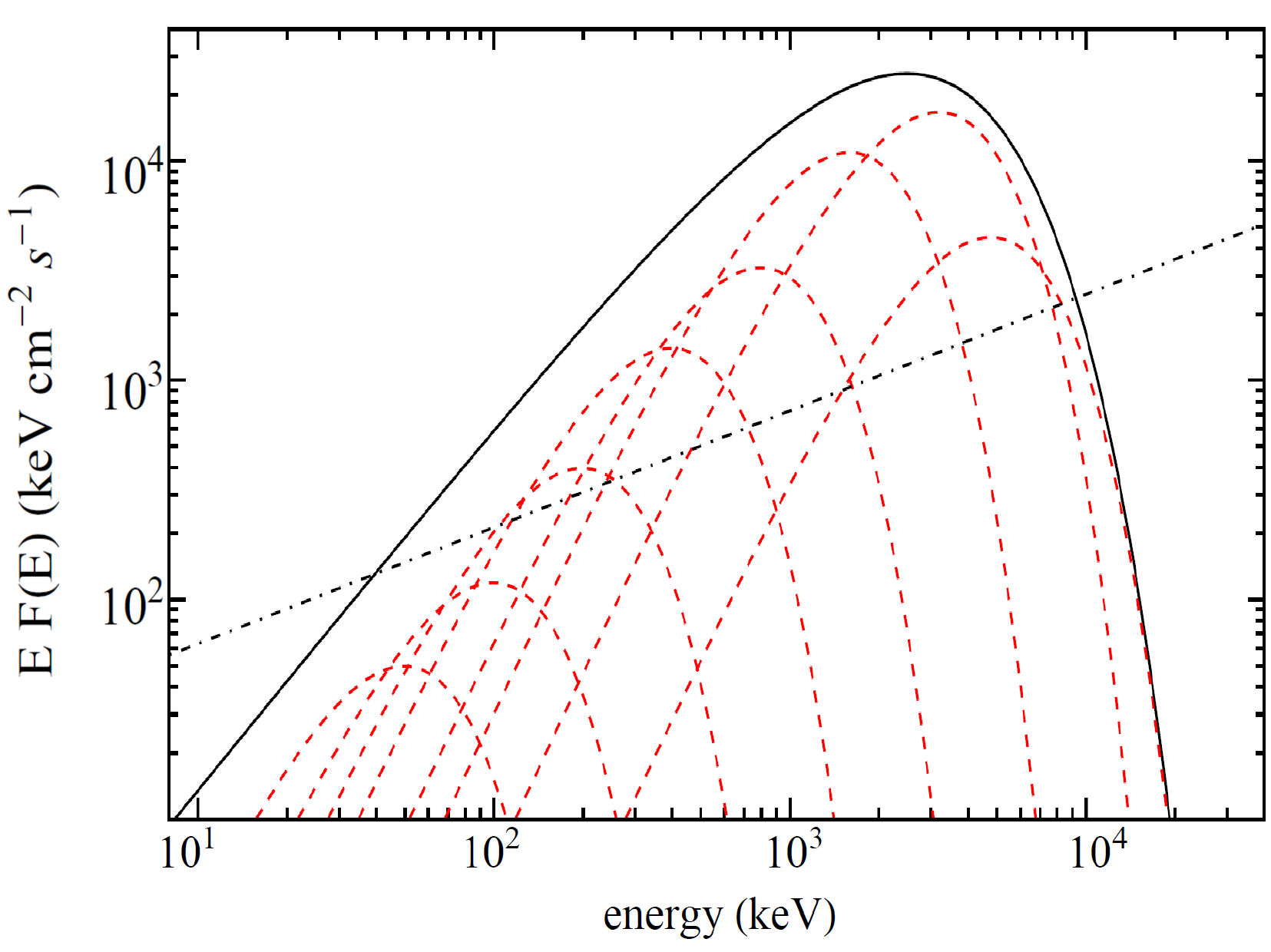}
    \caption{GRB 090510 spectrum of the Fermi-GBM NaI-n6, n7 and the BGO-b1 detectors, in the time interval T0 + $0.528$ to T0 + $0.644$ s (i.e., $0.275$--$0.338$ s rest-frame time interval). The best fit is a power-law model plus a Comptonized spectrum (black curve). The latter is viewed as a convolution of thermal components (dashed red curves), which are explained by repetitive transparencies of the $e^+e^-$ plasma produced by an overcritical electric field (see text for details). Figure adapted from \cite{2016ApJ...831..178R} with the authors' permission.}
    \label{fig:UPEbb}
\end{figure}

{We recall that the UPE phase comprises the emission within the GRB prompt, in the MeV energy regime, which is produced by the transparency events of the pair plasma. In GRB 090510, it occurs in the observer frame from T0 + $0.528$ to T0 + $0.644$ s, which corresponds to the $0.275$--$0.338$ s rest-frame time interval (see Fig. \ref{fig:UPEbb}).}

Previous analyses of the UPE phase, both in long and short GRBs, explain it from the transparency of the $e^+e^-$-photon plasma, expected to be created by QED vacuum polarization, expanding to ultrarelativistic velocity in a poorly baryon-contaminated ambient, reaching high Lorentz factor (see, e.g., \cite{1999A&A...350..334R,2000A&A...359..855R,2021PhRvD.104f3043M,2022EPJC...82..778R}, and references therein). 

The idea that $e^+e^-$ pairs from QED vacuum polarization could play a relevant role in GRBs dates back to the \citet{1975PhRvL..35..463D} analysis using the Kerr-Newman BH solution \cite{1965JMP.....6..918N}. The systematic application of this idea started to grow with the concept of \textit{dyadosphere} by \citet{1998A&A...338L..87P}, the region where pair formation from QED vacuum breakdown takes place outside a charged Reissner-Nordstr\"om BH \cite{1916AnP...355..106R,1918KNAB...20.1238N}. In that case, no magnetic fields are present, so the BH produces pairs if its charge is above the value $Q = E_c r_+^2$, where $E_c = m_e^2 c^3/(e \hbar)$ is the critical electric field for pair creation from vacuum, and $r_+$ is the BH event horizon. These conditions are therefore expected for charge-to-mass ratios $Q/(\sqrt{G}M) \gtrsim 4 M E_c G^{3/2}/c^4 \approx 7.5\times 10^{-6} (M/M_\odot)$, i.e., $Q \gtrsim 8\times 10^{33} e (M/M_\odot)^2 = 3.85\times 10^{24}(M/M_\odot)^2$ esu.

In \cite{1998A&A...338L..87P}, the pair density, the number of pairs, average energy, and the plasma temperature were estimated, evolving into the \textit{fireshell} model of GRBs \cite{1999A&A...350..334R,2000A&A...359..855R}. The thermodynamic properties of such a pair plasma determine the subsequent plasma dynamics and transparency that explain the GRB prompt emission \cite{1999A&A...350..334R,2000A&A...359..855R,2021PhRvD.104f3043M,2022EPJC...82..778R}. 
The concept of \textit{dyadoregion} covers additional geometries; for example, the dyadoregion of the Kerr-Newman solution can have a torus-like shape \cite{2009PhRvD..79l4002C}.

One of the main results of this model is that the transparency of the pair plasma leads to a characteristic of a blackbody component at MeV energies in the prompt emission. More recently, UPE analyses of long GRBs unveiled the presence of a \textit{hierarchical} (\textit{self-similar}) structure of the UPE, e.g., in GRB 190114C \cite{2021PhRvD.104f3043M} and GRB 180720B \cite{2022EPJC...82..778R}. The time-resolved UPE spectra of these GRBs, i.e., their UPE spectra in rebinned time intervals, are fitted by analogous composite blackbody+cutoff powerlaw spectra. This UPE observational feature has been interpreted as the repetitive occurrence of a microphysical phenomenon on ever shorter timescales, a succession of transparencies of the pair plasma.

The GRB 090510 UPE is shown in Fig.~\ref{fig:lc} from $0.25$ s to about $0.5$ s in the rest-frame \cite{2016ApJ...831..178R}. There, we individuated the UPE of this source, indicated as peak-GRB (P-GRB), and here reproduced in Fig. \ref{fig:UPEbb}. This figure shows that the UPE integrated spectrum of GRB 090510 can be interpreted as a convolution of blackbodies. This result suggests that for GRB 090510, there is an analogous hierarchical structure in the UPE.

Therefore, we also adopt here that QED pair formation from an overcritical electric field originates the pair plasma, whose expansion and transparency explain the UPE of GRB 090510. However, we introduce a new UPE's view that represents a paradigm shift. We here interpret the UPE via the $e^+e^-$ pair plasma formed by the electromagnetic field configuration around the merged object before the BH formation, which we model as that of a charged, rotating magnetic dipole obtained in \citet{1973ApL....13..109R} (see \ref{app:A} for details). The two approaches yield comparable quantitative results because, in both cases, the intensity of the electric field is constrained to be at least $E_c$ for it to create pairs. As shown in \ref{app:A}, the charge that minimizes the electromagnetic energy of the charged rotating magnetic dipole, and the \textit{effective charge} associated with the Faraday-induced electric field, depend on the product $J B_p$, and lead to charge-to-mass ratios of a few $10^{-5}$. This value is indeed similar to the BH charge of the aforementioned dyadospheres\footnote{Analogous considerations have been applied in the case of long GRBs in the framework of the Wald solution (see, e.g., \cite{2019ApJ...886...82R,2020EPJC...80..300R,2021A&A...649A..75M,2021PhRvD.104f3043M,2022EPJC...82..778R,2022ApJ...929...56R}).}. It is worth noticing that, from a quantitative viewpoint, the case of a rotating magnetic dipole without charge, in which there is only effective charge associated with the electric field given by Faraday induction, leads to similar results as the one considered here in the presence of a charge that minimizes the energy of the electromagnetic field.

Having clarified the key conceptual and qualitative differences between the present and previous UPE interpretations, we proceed to the quantitative analysis in the present approach. The BH birth is individuated by the onset of the GeV emission {(see Section \ref{sec:6})}, which occurs at $t_{\rm BH}\approx 0.5$ s in the rest frame. Since the UPE phase occurs before the GeV emission (see Fig. \ref{fig:lc}), i.e., before the BH forms, it must be explained via the massive metastable NS and the strong magnetic field, while the NS approaches the critical mass point. Thus, we can still use the electromagnetic field structure of a rotating magnetic dipole in this analysis (see \ref{app:A}).

The spontaneous $e^+e^-$ pair creation in the exterior of the merged object occurs when and where the induced electric field (in the frame where it is parallel to the magnetic field) reaches overcritical values. In \ref{app:A}, we have determined the associated dyadoregion radius $r_d(\theta)$ and width $\Delta_d$ (see Eqs. \ref{eq:rdya}), as well as the electromagnetic energy (Eq. \ref{eq:Epairs}). These properties depend on the value of $B_p$, $R$, and $\Omega$.

To estimate the angular velocity, we analyze the evolution of the merged object. We recall that the measurements of the mass of PSR J0952--0607 constrain the critical mass of a non-rotating NS to be $M^{J=0}_{\rm crit}\gtrsim 2.35 M_\odot$ \cite{2022ApJ...934L..17R}. Since the total mass of the binary is larger than this value and the rotation is high at coalescence, the merged core will likely have a mass larger than $M^{J=0}_{\rm crit}$. This core reaches uniform rotation on a millisecond timescale \cite{2025PhRvD.111d4038B}, becoming axially symmetric, GW emission ceases, and the object crosses the mass-shedding, Keplerian limit to enter into the supramassive, metastable regime. The object evolves in the metastable region, closely following a sequence of constant baryonic mass but losing angular momentum, ultimately reaching a critical point of gravitational instability. The latter is set by the turning points of the constant angular momentum sequences. At the critical mass point, it will have an angular momentum $0<J_{\rm crit}<J_K$, being $J_K \approx 0.7 G M^2/c$, the angular momentum of an NS at the Keplerian limit. The above value of $J_K$ is nearly independent of the nuclear equation of state of NS matter \cite{2015PhRvD..92b3007C}. Since $J = I \Omega$, the angular velocity at the Keplerian limit is $\Omega_K = 0.7 G M^2/(c I)$. Hence, assuming energy and angular momentum conservation during the collapse, the newborn BH will have $J < 0.7 G M^2/c$, so a spin parameter $\alpha \equiv cJ/(G M^2) < 0.7$. BHs formed by the prompt collapse of the merged object can have initial larger values of the dimensionless spin. The above scenario is consistent with NS-NS merger numerical relativity simulations (see, e.g., \cite{2006PhRvL..96c1101D,2006PhRvL..96c1102S,2006PhRvD..73j4015D,2007CQGra..24S.207S,2008PhRvD..77d4001S,2011ApJ...732L...6R}).

{In Section \ref{sec:6}}, from the request that the BH extractable energy powers the observed GeV emission, we obtain the BH mass $M=2.36 M_\odot$ and dimensionless spin $\alpha = 0.22$. From energy and angular momentum conservation, the above suggests the angular velocity of the merged core at the critical point was $\Omega_{\rm crit} = J_{\rm crit}/I \approx 0.22 G M^2/(c I) \approx 4.05\times 10^3$ rad s$^{-1}$, where we have used the moment of inertia $I = (2/5) M R^2 = 2.75\times 10^{45}$ g cm$^2$ (which agrees with numerical results; see, e.g., \cite{2015PhRvD..92b3007C}). The fact that $\Omega_{\rm crit} < \Omega_K$ agrees with our picture that the BH was formed via a fast but delayed collapse, triggered when the merged object reached the instability by angular momentum losses.

According to our picture, the pair creation should end by $t_{\rm BH}\approx 0.5$ s, i.e., the start of the GeV emission. This means we must request that the condition for pair creation is no longer fulfilled at that moment. This implies that, when $\Omega = \Omega_{\rm crit}$, $\Delta_d \to 0$, or equivalently, $\tilde E(R) \to E_c$ (while at times $t < t_{\rm BH}$ is overcritical). The magnetic field consistent with this request is given by Eq. (\ref{eq:Bmin}) in \ref{app:A}, which for the above critical angular velocity and fiducial radius leads to
\begin{equation}\label{eq:Bminvalue}
    B_{p, \rm min} = \frac{3}{4}\left(\frac{\Omega_{\rm crit} R}{c}\right)^{-1} = 2.04 \times 10^{14}\,\,\text{G}.
\end{equation}

{Having set the value of the magnetic field, the radius and magnetic field, we can discuss the value of the charge that minimizes the electromagnetic field energy (see \ref{app:A} for details). For a homogeneous sphere ($k=2/5$), and the mass and radius used in the GRB 090510 analysis, $M=2.4 M_\odot$ and $R = 12$ km, so ${\cal C} = 0.29$, we obtain from Eq. (\ref{eq:QwithI}), $|Q| = 2.82 (c^3/G) J B_p$, so $|Q_{\rm tot}(\theta=0)| =11.28 (c^3/G) J B_p$. This charge implies a charge-to-mass ratio $|Q|/(\sqrt{G}M) = 2.82 (c^3/G^{3/2}) a B_p$, where $a \equiv J/M$ is the specific angular momentum. For the above magnetic field, $B_p = 2\times 10^{14}$ G, we have $|Q|/(\sqrt{G}M) = 1.3\times 10^{-5}$, and $|Q_{\rm tot}(\theta=0)|/(\sqrt{G}M) = 5.2\times 10^{-5}$. Interestingly, the effective charge associated with the Faraday-induced electric field is, on the pole, three times the above charge (see Eq. \ref{eq:sigma}). These charge and effective charge values are, indeed, comparable with  $Q/(\sqrt{G}M)  = r_+^2 E_c/(\sqrt{G}M) \approx 4 M E_c G^{3/2}/c^4 = 1.8\times 10^{-6}$, the charge-to-mass ratio that a Reissner-Nordstr\"om BH of the same mass needs to have a dyadosphere.
}

Thus, at times $t \leq t_{\rm BH}$, $\Omega$ decreases until $\Omega_{\rm crit}$, and assuming the magnetic field is the same, the dyadoregion size decreases and the energy available for the pairs, which is given by the electromagnetic energy within the dyadoregion, decreases, until it becomes zero at $t = t_{\rm BH}$. It is this evolution of the pair plasma initial conditions that explains the different transparencies observed (different blackbodies).

The energy available for the pairs is given by Eq. (\ref{eq:Epairs}). Assuming that during the UPE the magnetic field holds at a constant value given by Eq. (\ref{eq:Bminvalue}), the available energy at the beginning of the evolution of the merged object in the supramassive metastable region, i.e., for $\Omega = \Omega_K = 0.7 G M^2/(c I) = 1.29\times 10^4$ rad s$^{-1}$, is ${\cal E}_{e^+e^-} = 7.92\times 10^{44}$ erg. The width of the dyadoregion (along the polar axis) is $\Delta_d (\theta=0) \approx 0.41 R \approx 4.92$ km. The integrated energy of the UPE is $E_{\rm UPE} = 3.95\times 10^{52}$, so applying a beaming-factor correction of $f_b\approx 1/3$ (see \ref{app:A}), $E'_{\rm UPE} = f_b E_{\rm UPE} =  1.32\times 10^{52}$ erg, we estimate that the UPE consists of $N \approx E_{\rm UPE}/{\cal E}_{e^+e^-} \approx 2\times 10^7$ impulses.    

We now estimate the physical parameters of the plasma successive transparencies using the blackbodies inferred from the spectral analysis of the UPE of GRB 090510 in \cite{2016ApJ...831..178R}. For this task, we follow an analogous analysis to the one performed in  \cite{2021PhRvD.104f3043M,2022EPJC...82..778R}, approximating the dynamics to that of a spherically symmetric ultrarelativistic self-accelerated $e^+e^-$-photon plasma in a poorly baryon-contaminated medium. Under this assumption, the Lorentz factor at transparency, the baryon load parameter, and the transparency radius are well approximated by (see \cite{1999A&A...350..334R,2000A&A...359..855R,2021PhRvD.104f3043M,2022EPJC...82..778R}, for details)
%
%
\begin{equation}
    \Gamma \approx \left( \frac{a T_{\rm obs}^4 \sigma_T \Delta_d}{16 m_N c^2}\frac{1-\epsilon_{\rm BB}}{\epsilon_{\rm BB}} \right)^{1/3},\quad
    {\cal B} \equiv \frac{M_B c^2}{{\cal E}_{e^+e^-}} = \frac{1-\epsilon_{\rm BB}}{\Gamma-1}\quad
    R_{\rm tr} = \sqrt{\frac{\sigma_T}{8\pi}\frac{{\cal B}{\cal E}_{e^+e^-}}{m_N c^2}},
\end{equation}
where $M_B$ is the mass of the baryonic matter loaded into the plasma, $\epsilon_{\rm BB} \equiv E_{\rm BB}/{\cal E}_{e^+e^-}$, $m_N$ is the nucleon mass, $\sigma_T$ is the Thomson cross-section, and $a = 4\sigma/c$, being $\sigma$ the Stefan-Boltzmann constant. Table \ref{tab:UPE090510} shows the inferred model parameters\footnote{These parameters are comparable to those obtained in the UPE analysis of the long GRBs 190114C \cite{2021PhRvD.104f3043M} and 180720B \cite{2022EPJC...82..778R}, in the context of an $e^+e^-$ plasma formed around a Kerr BH in an external magnetic field. The reason for this result is that the size and energy of the dyadoregion of the present case of an NS approaching the critical mass are quantitatively similar to the corresponding values for a Kerr BH in an external magnetic field with comparable values of mass and magnetic field strength.}.

\begin{table}
    \centering
    \begin{tabular}{c|c|c|c|c}
    $i$-th event & $k_B T_{{\rm obs},i}$ (keV) & $\Gamma_i$ & ${\cal B}_i$ ($10^{-2}$) & $R_{{\rm tr},i}$ ($10^{10}$ cm)\\
     \hline 
     1 & 1216 & 1785.61 & 0.032 & 0.21 \\
 2 & 811 & 1040.49 & 0.056 & 0.28 \\
 3 & 405 & 412.24 & 0.14 & 0.44 \\
 4 & 203 & 164.14 & 0.36 & 0.70 \\
 5 & 101 & 64.71 & 0.91 & 1.13 \\
 6 & 51 & 26.02 & 2.32 & 1.80 \\
 7 & 25 & 10.06 & 6.40 & 2.99 \\
 8 & 13 & 4.21 & 18.09 & 5.02\\
 \hline 
    \end{tabular}
    \caption{Inferred properties of the $e^+e^-$ pair plasma transparencies observed in the UPE of GRB 090510 (see Fig. \ref{fig:UPEbb}). The value of $k_B T_{{\rm obs},i}$ and $\epsilon_{{\rm BB},i}$ are obtained from the analysis in \cite{2016ApJ...831..178R}. We estimate the latter as $\epsilon_{{\rm BB},i} = E_{{\rm BB},i}/{\cal E}_{e^+e^-,i} \approx E_{\rm BB}/E_{\rm UPE} \approx 0.42$ \cite{2016ApJ...831..178R}.}
    \label{tab:UPE090510}
\end{table}

In the above UPE analysis, we have used the minimum magnetic field value given by Eq. (\ref{eq:Bminvalue}), which is of a few $10^{14}$ G, a value smaller than the magnetic field inferred from the precursor, which is of a few $10^{16}$ G (see section \ref{sec:3}). We can hypothesize (at least) three physical situations: 

(i) The magnetic field strength is constant and has the value used in the UPE. In this case, the explanation of the precursor luminosity and energetics would require a very tiny beaming angle of the order of $0.1^\circ$. 

(ii) The magnetic field strength is constant and has the value inferred from the precursor. In this case, the qualitative picture of the UPE is the same, and only the quantitative conclusions would change, e.g., the energy available for pairs, which scales as $B_p^{7/4}$ (see Eq. \ref{eq:Epairs}), hence the number of impulses. However, since the request of transparency of the GeV emission (see Section \ref{sec:6}) requires the magnetic field to be $\lesssim 10^{11}$ G, the magnetic field must suffer a sudden, very large drop of five orders of magnitude as the BH forms. This drop can be caused by counter-rotating currents created by the motion of the pairs (e.g., \cite{2021PhLB..82036562C}), or the lack of anchoring by the BH horizon (see, e.g., \cite{2021PhRvL.127e5101B}, for numerical simulations), which implies the magnetic field strength is limited by the hydromagnetic equilibrium of the surrounding accretion disk. 

(iii) The magnetic field decreases from the precursor to the UPE. The field decay could be the result of the release of magnetic energy in the precursor, e.g., by magnetic reconnection, which can cause a rearrangement of the magnetic field structure (see, \cite{2022LRSP...19....1P}, for a recent review on the subject), as in solar corona flares (e.g., \cite{2020Sci...367..278F}) or pulsar flares (e.g., \cite{2005A&A...442..579C}). 

Independent of what the precise situation is, we can conclude from the analysis of the precursor and the UPE that the magnetic field around the merged object must be strong, in the range $\sim 10^{14}$--$10^{16}$ G.

\section{Post-merger spikes}\label{sec:5}

\begin{figure}
    \centering
    \includegraphics[width=0.75\hsize,clip]{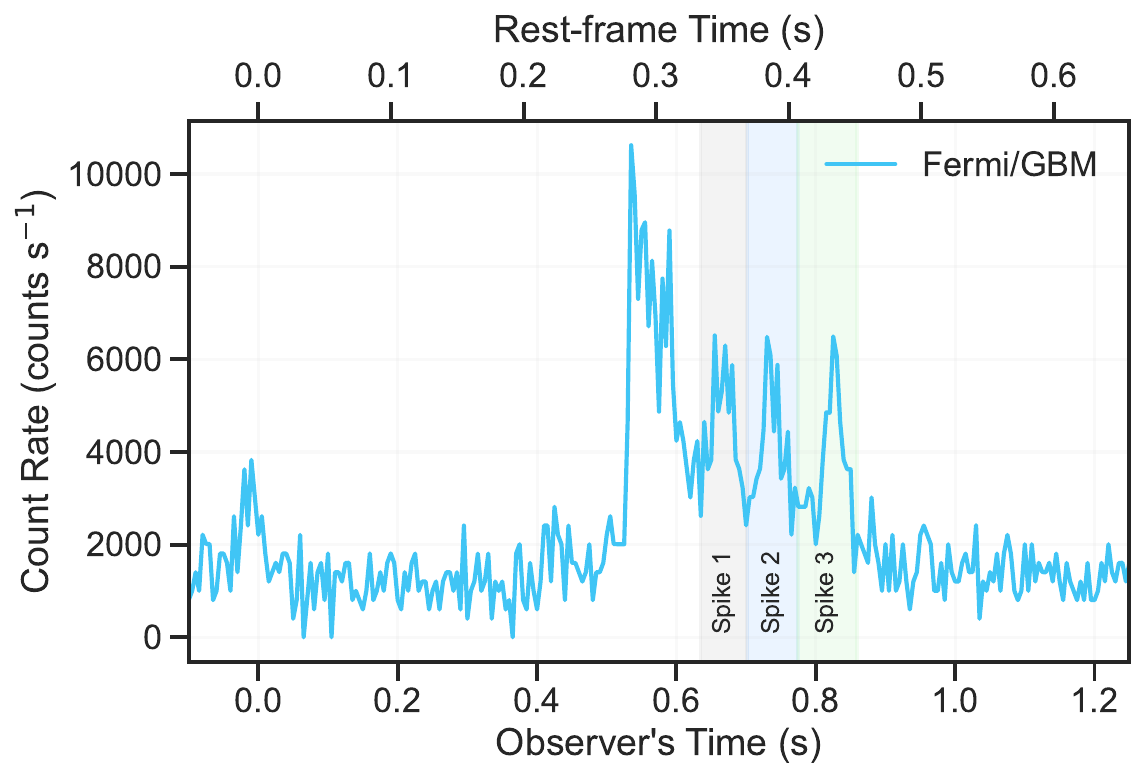}
    \caption{Light curve of the three spikes following the main pulse of the prompt emission observed by Fermi-GBM (NaI-n6). The peaks occurs at a rest-frame $t_{\rm sp1}= 0.353 \pm 0.005$ s, $t_{\rm sp2} = 0.388 \pm 0.005$ s, and $t_{\rm sp3} = 0.429 \pm 0.008$ s. The separation between successive peaks is $\Delta_{\rm sp} \approx 40$ ms. The timescale between spikes could be explained by the timescale of Lense-Thirring precession frequency of the accreting matter onto the BH of mass $2.36 M_\odot$ with spin $\alpha=0.22$, at a radius $r = 1.18 R_{\rm lco} = 6.20 M = 21.60$ km.}
    \label{fig:spikes}
\end{figure}

Another interesting feature of GRB 090510 is that its light curve shows, after the main pulse (UPE) of the prompt emission, three spikes separated by nearly the same time interval (see Fig. \ref{fig:lc}). Spike 1 peaks at a rest-frame time $t_{\rm sp1}= 0.353 \pm 0.005$ s, has a duration $\Delta t_{\rm sp1} = 0.031$ s, and we estimate released $E_{\rm sp1}=(5.703 \pm 0.548)\times 10^{50}$ erg of energy. Spike 2 shows $t_{\rm sp2} = 0.388 \pm 0.005$ s, $\Delta t_{\rm sp2} = 0.036$ s, and $E_{\rm sp2}=(6.259 \pm  0.846)\times 10^{50}$ erg. Spike 3 is characterized by $t_{\rm sp3} = 0.429 \pm 0.008$ s, $\Delta t_{\rm sp3} = 0.045$ s, and $E_{\rm sp3}=(5.026 \pm  0.664)\times 10^{50}$ erg. As these spikes are observed before the GeV emission, in our picture, they must be related to the activity of the merged object as it reaches the critical mass.

The emission from accreting BHs and NSs in X-ray binaries shows quasi-periodic oscillations (QPOs), some of them of tens of Hz. These QPOs are typically explained via the characteristic frequencies associated with the motion of the accreting matter in the strong field regime, following the relativistic precession model \cite{1998ApJ...492L..59S,1999PhRvL..82...17S,1999ApJ...524L..63S}, some of them associated with the nodal precession frequency, i.e., the Lense-Thirring frequency. 

Thus, we advance and investigate the possibility that the afore-mentioned three spikes observed in GRB 090510 could be a result of the Lense-Thirring (frame-dragging) precession of the material surrounding the massive, fast-rotating NS near the critical mass. Generally, the modeling of the exterior spacetime of a rotating NS is obtained via the numerical solution of the Einstein equations in axial symmetry, with a matter content described by a realistic nuclear equation of state. It has been shown that as the NS approaches the critical mass, its multipole moments approach those of a Kerr BH (see, e.g., Fig. 13 in \cite{2015PhRvD..92b3007C}, and also references therein). Hence, its exterior spacetime and its associated properties approach those of the Kerr solution. This result is confirmed by the analysis of the last stable circular orbit (LCO) properties around rotating NSs \cite{2017PhRvD..96b4046C}. In particular, it can be seen from Figs. 3 and 4 therein, that the energy and angular momentum of the LCO of an NS of mass $\approx 2.4 M_\odot$ and spin $\alpha = cJ/(G M^2) = 0.2$, as the one we are analyzing here, approach those given by the Kerr solution with the same parameters. 

With this information in mind, we adopt the Kerr metric with these parameters in the present analysis. Thus, the Lense-Thirring angular frequency on the equatorial plane is
\begin{equation}\label{eq:OmegaLT}
    \Omega_{\rm LT} = -\frac{g_{t \phi}}{g_{\phi \phi}} = \frac{2 \hat a \hat M r}{A}c,
\end{equation}
where $A = (r^2+\hat a^2)^2 - \Delta \hat a^2$ and $\Delta = r^2 - 2 \hat M r + \hat a^2$, where $\hat M = G M/c^2$ and $\hat a = J/(M c) = \alpha\,G M/c^2$ are the mass and specific angular momentum in geometric units. For the above mass and spin, we obtain the frequency at the LCO, $r = R_{\rm LCO} = 5.27 M$, $\Omega_{\rm LT} = 255.67$ rad s$^{-1}$, to which corresponds a timescale $P_{\rm LT} = 2\pi/\Omega_{\rm LT} = 24.57$ ms. The timescale increases rapidly with distance from the central object. The time interval between the spikes of GRB 090510 is $\Delta t_{\rm sp}\approx 40$ ms. From the above model, we obtain $P_{\rm LT} = \Delta t_{\rm sp} = 40$ ms at $r = 1.28 R_{\rm LCO} = 6.20 M = 21.60$ km. Interestingly, these numbers are similar to those obtained for X-ray binaries, see, e.g., the case of XTE J1550--564, for which the QPO analysis led to $\alpha = 0.34$ and radius $r = 1.13\,R_{\rm LCO}$ \cite{2014MNRAS.439L..65M}. {It is interesting to note that the existence of QPOs in long GRBs has been suggested in several sources, e.g., in GRB 220711B, a $50$ s QPO in the X-ray afterglow has also been attributed to Lense-Thirring precession of a tilted disk \cite{2025ApJ...985...33G}. Because the timescale for the alignment of the BH and disk angular momentum should be longer than the total emission duration, the QPO timescale constrains the properties of the BH-disk system \cite{1975ApJ...195L..65B}, and the properties of the system formation (see, e.g., \cite{2025ApJ...985...33G}, for details).
}

The above numbers are very suggestive and are left for observational verification of QPOs in GRBs, which, due to the transient nature of the system, remains a challenging and formidable task (see \cite{2025A&A...697A.228G} for a recent discussion of the subject).

\section{The GeV emission}\label{sec:6}

{In our scenario, the GeV radiation is associated with emission originating from the newborn BH. The latter forms from a delayed collapse of the merged object, so the time interval from merger to the GeV emission, which is $t_{\rm BH} \approx 0.5$ s, indicates approximately the time in which the merged object spins down and reach the critical mass point, as described in Section \ref{sec:5}.}

In the GeV phase following the UPE, it is expected that the magnetic field has been reduced to values $\lesssim 10^{11}$ G, as requested by the transparency of the system to GeV photons \cite{2021A&A...649A..75M,2022ApJ...929...56R}. The magnetic field decrease could have been screened by the plasma (see, e.g., \cite{2021PhLB..82036562C}), or reduced by the inability of the BH and the surrounding material to hold strong magnetic fields \cite{2021PhRvL.127e5101B}.

{Using the Wald solution, it has been shown that} the electric field induced by such a magnetic field and the BH rotation is sufficient to accelerate electrons whose radiation explains the GeV emission \cite{2019ApJ...886...82R, 2020EPJC...80..300R, 2021A&A...649A..75M,2022ApJ...929...56R}. {The outward acceleration of electrons} occurs in a conical region with semi-aperture angle $\theta_{\rm GeV}\approx 60^\circ$ 
around the BH rotation axis \citep{2021A&A...649A..75M,2022ApJ...929...56R}. {It is thus expected that a return interaction with the BH mediates the energy extraction process, as a closure of energy and angular momentum conservation. Indeed, it has been shown that protons inside the polar conical region and some electrons around the equator fall into the BH, inject a net negative energy and angular momentum into it, acting as the energy and angular momentum extracted to the BH, while producing a small increase of $M_{\rm irr}$ (see \cite{2023EPJC...83..960R,2024EPJC...84.1166R} for details).} This scenario is very different from the previous interpretation of the GeV emission of GRB 090510 based on accretion onto the BH \cite{2016ApJ...831..178R}.

{Therefore,} the energy budget required for the GeV emission is $E_{\rm GeV}'\equiv E_{\rm GeV,iso} [1-\cos (\theta_{\rm GeV})] = E_{\rm GeV,iso}/2$, where $E_{\rm GeV,iso}$ is the measured isotropic energy released in the GeV energy band, which we now estimate. The GeV light curve (see Fig. \ref{fig:lc}) is fitted with a smooth broken power-law model \cite{1999A&A...352L..26B}: $L_{\rm GeV}(t) = A \left[\left(t/t_{\text{break}}\right)^{\alpha_1/\delta} + \left(t/t_{\text{break}}\right)^{\alpha_2/\delta} \right]^{\delta}$, where $A$ is the normalization at $ t_{\rm break}$, $ \alpha_1 $ and $ \alpha_2 $ are the temporal decay indices before and after the break, respectively, and $\delta$ controls the sharpness of the transition. The best-fit parameters of the luminosity observed by the Fermi-LAT in the $0.1$--$100$ GeV band are $A = (1.97 \pm 0.70) \times 10^{51}$ erg s$^{-1} $, pre-break index $ \alpha_1 = -1.194 \pm 0.086$, post-break index $ \alpha_2 = -2.593 \pm 0.209 $, break time $ t_{\text{break}} = 2.018 \pm 0.400$ s, and smoothness parameter $ \delta = 0.333 \pm 0.369$. With the above, we obtain $E_{\rm GeV,iso} = \int_{t_{\min}}^{t_{\max}} L_{\rm GeV}(t) \, dt=5.09 \times 10^{52}$ erg, where $t_{\min} \approx 0.5$ s and $t_{\rm max} \approx 2$ s (see Fig. \ref{fig:lc}). Notice that, since the luminosity decreases following a power-law function, the value of the integral is not sensitive to larger values of $t_{\rm max}$. The smooth transition between power-law segments ($\delta=0.33$) indicates a gradual rather than abrupt change in the emission properties. This favors scenarios where the break results from a continuous evolution of physical parameters rather than a discrete transition. Therefore, the required energy budget is $E_{\rm GeV}' = 2.54\times 10^{52}$ erg.

The energy reservoir of the GeV emission is the BH extractable energy, i.e., $E_{\rm ext} \equiv (M - M_{\rm irr}) c^2 = E_{\rm GeV}'$, which we can use to estimate the BH mass
\begin{equation}\label{eq:BHenergy}
M = M_{\rm irr} + E_{\rm GeV}'{c^2}.
\end{equation}
The BH parameters are related via the mass-energy formula \citep{1970PhRvL..25.1596C,1971PhRvD...4.3552C,1971PhRvL..26.1344H}, $M^2 = M_{\rm irr}^2 + c^2 J^2/(4 G^2 M_{\rm irr}^2)$, which we can use to estimate the BH spin
\begin{equation}\label{eq:Mirr}
    \alpha = 2 \frac{M_{\rm irr}}{M} \sqrt{1-\left( \frac{M_{\rm irr}}{M}  \right)^2}.
\end{equation}

In deriving Eq. (\ref{eq:BHenergy}), we have implicitly assumed that the BH irreducible mass remains constant during the emission process {of the energy $E'_{\rm GeV}$. This assumption agrees with the electrodynamical process of energy extraction adopted here and described above.}

Having clarified this point, we can estimate the BH mass and spin given $M_{\rm irr}$. For the latter, we adopt $M_{\rm irr}=2.35 M_\odot$, the lower limit to the critical mass of non-rotating NS given by the mass of PSR J0952--0607 \cite{2022ApJ...934L..17R}. With all the above, we estimate $M = 2.36 M_\odot$ and $\alpha = 0.22$.

In the modeling of the GeV emission, we have used the model of our previous studies \cite{2019ApJ...886...82R,2020EPJC...80..300R,2021A&A...649A..75M,2022ApJ...929...56R,2023EPJC...83..960R,2024EPJC...84.1166R} based on the Wald solution of a Kerr BH in an asymptotically uniform, aligned test magnetic field \cite{1974PhRvD..10.1680W}, which extends the solution by Hanni and Ruffini of a Schwarzschild BH in an asymptotically uniform, aligned test magnetic field \cite{1976NCimL..15..189H} (see Figs. 1 and 2 therein for the lines of force of the field). The magnetic field around the Kerr BH is $10^{11}$ G, so $B M \approx 10^{-11}$, which validates our treatment using the test magnetic field solution of the Einstein-Maxwell equations. A recent paper by Podolsky and Ovcharenko \cite{2025arXiv250705199P} presents a new exact solution of the Einstein-Maxwell equations for a Kerr BH in an external asymptotically uniform magnetic field. The algebraic type of the solution is different than the one of the above approximate solutions, so it is not possible to perform a one-to-one comparison of them, even in the weak-field (test) regime of the magnetic field, although it remains as an interesting task for the future to analyze the possible astrophysical consequences of new exact solutions.

\section{Late X-ray emission}\label{sec:7}

The material that remained bound circularizes around the newborn BH and accretes onto it. The luminosity release in the accretion process powers the X-ray emission. The evolution of the BH mass and spin in this process can be obtained from the energy and angular momentum conservation equations
\begin{equation}\label{eq:malphadot}
        \dot M = \epsilon\,\dot m,\quad
        \dot \alpha = \left(\frac{l}{\epsilon} - 2 \alpha\right)\,\frac{\dot M}{M},
\end{equation}
where $\epsilon$ and $l$ are the energy per unit mass and dimensionless angular momentum (i.e., specific angular momentum, per unit BH mass) of the accreted material, being $\dot m$ the accretion rate. The gravitational energy gain of the inflowing matter until the inner accretion radius powers the luminosity, so $L_{\rm acc} = (1-\epsilon)\dot m c^2$, which we can use to rewrite Eq. (\ref{eq:malphadot}) as
\begin{equation}\label{eq:Mdot2}
    \dot M = \frac{\epsilon}{1-\epsilon} \frac{L_{\rm acc}}{c^2}.
\end{equation}
For $\epsilon$ and $l$, we adopt the values of the (co-rotating) last circular orbit (LCO), of radius $R_{\rm lco}$, which are all analytic functions of $\alpha$ \cite{1971ESRSP..52...45R,1972ApJ...178..347B,1974bhgw.book.....R}. Thus, given a luminosity $L_{\rm acc} = L_X(t)$, we can integrate (numerically) the system of differential equations given by Eqs. (\ref{eq:malphadot}) and (\ref{eq:Mdot2}) and obtain the BH evolution in this post-merger phase. 

The X-ray luminosity of the afterglow can be also fitted with a broken power-law function (see Fig. \ref{fig:lc}) with parameters $A = (5.09 \pm 0.86)\times 10^{47}$ erg s$^{-1}$, $\alpha_1 = -0.567 \pm 0.071$, $\alpha_2 = -1.992 \pm 0.114$, $t_{\rm break} = 638.466 \pm 84.405$ s, and $\delta = -0.050 \pm 0.384$. This leads to an energy release in X-rays $E_X = 8.23\times 10^{50}$ erg. With this luminosity, we integrate the evolution equations and obtain the BH evolution in the X-ray afterglow phase. {From the above, we obtain the maximum accretion rate, which corresponds to the rate at the beginning of the X-ray afterglow phase, $\dot m = 2.51\times 10^{-4} M_\odot$ s$^{-1}$. While a specific accretion disk modeling is beyond our present scope, it is worth mentioning that at these accretion rates photon trapping is strong, but also neutrino and antineutrino emission and annihilation (see, e.g., \cite{2016PhRvD..93l3004L,2016ApJ...833..107B,2018ApJ...852..120B,2021Univ....7....7U}), which could lead to X-ray emission (see, e.g., \cite{1987A&A...175..309B}).}

The accumulated accreted mass onto the BH is
\begin{equation}
    m_{\rm acc} = \int \frac{\dot m}{\epsilon} dt = 0.0089 M_\odot,
\end{equation}
where we have carried out the integral from the BH formation time (so assuming valid the extrapolation of the luminosity backward to $t=0.5$ s), to infinity. This mass value provides an estimate of the accretion disk mass. The accretion energy release increases to $1.042\times 10^{51}$ erg. The final BH mass and spin are $2.37 M_\odot$ and $\alpha = 0.23$.

The small mass inferred for the accretion disk is consistent with our assumption that the pre-merger system was (or was very close to) a mass-symmetric binary. For a mass ratio $q\equiv m_2/m_1 <1$, the tidal disruption radius is $\sim R_2\,q^{-1/3} > R_2$, so it would cause the primary to disrupt the secondary before merger. Under these conditions, the system would be composed of an NS surrounded by a massive accretion disk, instead of a BH surrounded by little mass, as our analysis suggests.


\section{Inferences on the GW emission}\label{sec:8}

From the inferred merging masses, we can infer the maximum energy released in GWs in the inspiral phase, as the difference in binary gravitational binding between an infinite separation and the merger distance, $r_{\rm mgr} = 2 R = 24$ km, i.e. $\Delta E_{\rm GW, insp} < G m_1 m_2/(2 r_{\rm mgr})\approx 0.018 M c^2 = 7.60\times 10^{52}$ erg. 

We can estimate the GW energy release during the merger from angular momentum conservation. The latter, from the merger to the BH+disk system formation requires
\begin{equation}\label{eq:Jconservation}
    J_{\rm bin, mgr} = J_{\rm BH} + J_d + \Delta J_{\rm rad},
\end{equation}
where $J_{\rm bin, mgr} = \mu \sqrt{G M_{\rm bin} r_{\rm mgr}}$ is the binary angular momentum at the merger point $r=r_{\rm mgr}$, $J_{\rm BH} = \alpha G M^2/c$ is the BH angular momentum, $J_d \approx l(\alpha) G M m_d/c$ is the approximate angular momentum of the disk, and $\Delta J_{\rm rad}$ is the angular momentum radiated away to infinity during the merger process. From Eq. (\ref{eq:Jconservation}), we can estimate the radiated angular momentum $\Delta J_{\rm rad} = 0.302 G M^2/c$. This angular momentum is mainly carried out by GWs, so $\Delta J_{\rm rad} = \Delta J_{\rm GW}$. Thus, the associated GW energy release is $\Delta E_{\rm GW,mgr} \sim \Omega_{\rm mgr} \Delta J_{\rm GW} \approx 0.007 M c^2 = 2.96\times 10^{52}$ erg, where we have used $\Omega_{\rm mgr} \approx \Omega_K \approx 0.023 c^3/(G M)$, which corresponds to $\approx 2$ kHz. 

Interestingly, the above inferred values of the disk mass and angular momentum, as well as the radiated energy and angular momentum during the merger phase agree with results from numerical relativity simulations of NS-NS mergers leading to prompt collapse into a BH (see, e.g., \cite{2018PhRvL.120k1101Z,2018ApJ...852L..29R}). 

These numbers imply that neglecting the energy and angular momentum radiated during the merger phase leads to an error of only $\sim 1\%$ in our estimate of the binary total mass, hence of the merging components. The pre-merger binary mass-energy is $M_{\rm bin} = m_1 + m_2$, where $m_{1,2}$ are the NS masses. Hence, we can safely approximate $M_{\rm bin} \approx M$, and assuming an equal-mass binary ($m_1 = m_2 = m$), $M_{\rm bin} = 2 m$, we obtain $m \approx M/2 \approx 1.2 M_\odot$. Interestingly, we obtain a mass close to the canonical mass and the average mass of the NS components of observed NS-NS binaries. We recall that our analysis has assumed the BH irreducible mass remains constant during the GeV emission. An increase of $M_{\rm irr}$ would reduce the extractable energy, leading to a higher value of the BH mass and spin. In turn, this would result in larger merging NS masses.

\section{Discussion and conclusion}\label{sec:9}

%
\begin{table}
    \centering
    \begin{tabular}{c|c}
    \hline $E_{\rm prec}$ & $2.28\times 10^{51}$ erg\\
    $E_{\rm UPE}$ & $3.95\times 10^{52}$ erg\\
    $E_{\rm GeV}$ & $5.09\times 10^{52}$ erg\\
    $E_X$ & $8.23\times 10^{50}$ erg\\
    \hline
         $M$ & $2.36 M_\odot$ \\
         $\alpha$ & $0.22$\\
         $M_{\rm irr}$ & $2.35 M_\odot$\\
         \hline
         $m_1 = m_2$ & $1.19 M_\odot$\\
         $B_p$ & $\gtrsim2.04\times 10^{14}$ G\\
         $\Delta E_{\rm GW, mgr}$ & $\sim 0.007 M c^2=2.96\times 10^{52}$ erg\\
         $\Delta J_{\rm GW, mgr}$ & $0.30\,G M^2/c$\\
         $m_d = m_{\rm acc}$ & $0.0089 M_\odot$\\
         \hline
    \end{tabular}
    \caption{Upper: GRB 090510 energetics, in particular the isotropic energy released in the precursor, UPE, and GeV emission. Middle: inferred parameters of the newborn Kerr BH (mass, spin, and irreducible mass). Lower: additional properties of the system inferred from the analysis, the mass of the merging NSs ($m_1$ and $m_2$), the strength of the magnetic field surrounding the merged object ($B_p$), the energy ($\Delta E_{\rm GW, mgr}$) and angular momentum ($\Delta J_{\rm GW, mgr}$) radiated in GWs during merger, and the disk mass (accreted) ($m_d = m_{\rm acc}$) by the BH in the post-merger phase.}\label{tab:BHparameters}
\end{table}

Table \ref{tab:BHparameters} summarizes the main inferred parameters of our NS-NS merger model of GRB 0905010. The progenitor is a binary composed of two NSs of $\approx 1.2 M_\odot$ each. The merger
formed a Kerr BH of mass $M = 2.36 M_\odot$ and angular momentum $J = 0.22 G M^2/c$. During the merger phase, magnetic energy of an amplified magnetic field (of up to a few $10^{16}$ G) powers the precursor of observed isotropic energy $E_{\rm prec} = 2.3\times 10^{51}$ erg. 

The electric field around the merged object polarizes the vacuum, leading to an $e^+e^-$ plasma. The latter self-accelerates, and due to the low baryonic matter pollution produced in the merger, reaches ultrarelativistic speeds. The plasma transparency leads to the UPE, releasing $E_{\rm UPE}= 3.95\times 10^{52}$ erg in MeV photons. We have derived the physical parameter of the plasma transparencies (see Table \ref{tab:UPE090510}). We have also clarified the two paradigm shifts in the origin of the pair plasma, related to our previous considerations. The overcritical electric field that triggers the spontaneous pair creation is not produced by a BH endowed with electric charge; it is induced by the rotating magnetic field around the merged object, before it forms a rotating Kerr BH by reaching the critical mass for gravitational collapse.

Some material remains bound to the system in a surrounding disk around the central remnant. We have shown that the Lense-Thirring precession of the material around the massive NS approaching the critical mass can produce QPOs that explain the three spikes observed after the main peak of the prompt emission, separated by time intervals of $40$ ms. 

The start of the GeV emission marks the formation of the Kerr BH. The induced electric field in this phase is undercritical (and the magnetic field $\lesssim 10^{11}$ G) but sufficient to accelerate electrons leading to GeV photons emitted in a confined region of semi-aperture angle of $\approx 60^\circ$ around the BH rotation axis, releasing the observed $E_{\rm GeV} = 5.09\times 10^{52}$ erg detected by Fermi-LAT. The extractable energy of the BH is the energy reservoir of the GeV emission. This scenario differs from the previous one in \cite{2016ApJ...831..178R} in terms of accretion power. The decrease of the magnetic field from the UPE to the GeV could be due to counter-rotating currents of the UPE pairs \cite{2021PhLB..82036562C}, the inability of the BH to anchor the magnetic field \cite{2021PhRvL.127e5101B}, and the surrounding accreting material that cannot sustain ultrastrong fields.

Because of angular momentum conservation, a small-mass accretion disk of $0.009 M_\odot$ forms around the BH. The accretion power explains the X-ray afterglow of $E_{X} = 8.23\times 10^{50}$ erg. Interestingly, our inferred low disk mass agrees with the values obtained in numerical relativity simulations of NS-NS mergers leading to BH formation.

Based on energy and angular momentum conservation, we have inferred that the merger could have released $\sim 0.007 M c^2 = 2.96\times 10^{52}$ erg of energy in GWs, most before BH formation. This energy release is at best comparable to the electromagnetic energy release (see Table \ref{tab:BHparameters}). While the electromagnetic emission from an ultramagnetized NS-NS merger is directly probed by the present short GRB data, it remains to probe the GW emission in these systems where the electrodynamics and associated emission by electric and magnetic fields well above $\sim 10^{14}$ G could have a non-negligible role in the dynamics of the merging binary (see, e.g., \cite{2022ApJ...940...90C,2023PhRvD.108b4020H,2024PhRvD.109h4048H,2024RAA....24k5002T}).

\begin{figure}
    \centering
    \includegraphics[width=0.7\hsize,clip]{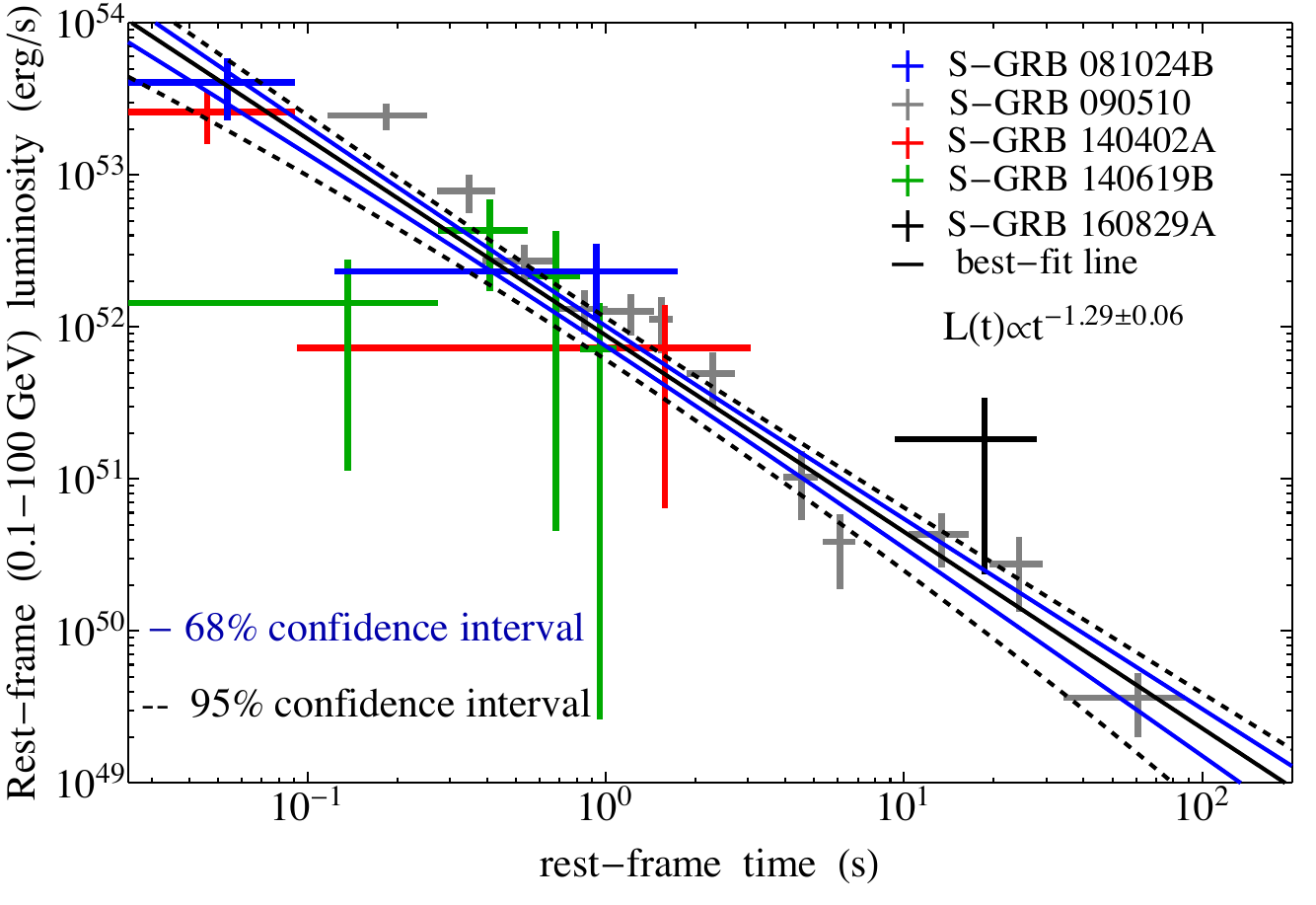}
    \caption{Rest-frame $0.1$--$100$ GeV isotropic luminosity of S-GRBs observed by Fermi-LAT. The black line indicates the common power-law behavior of the GeV emission with the slope $-1.29 \pm 0.06$. Reproduced from \cite{2018arXiv180207552R} with the authors' permission.}
    \label{fig:sGRBsFermiLAT}
\end{figure}

The above NS-NS merger scenario must not be unique to GRB 090510. Short GRBs emitting $E_{\rm iso}\gtrsim 10^{52}$ erg have been classified as authentic short GRBs (S-GRBs) \cite{2016ApJ...832..136R} and, in analogy to the long GRBs with $E_{\rm iso}\gtrsim 10^{52}$ erg, the BdHN I (see \cite{2023ApJ...955...93A}, and references therein), S-GRBs are thought to lead to a BH central remnant and an associated GeV emission. In this line, short GRBs with $E_{\rm iso} \lesssim 10^{52}$ erg, classified as short gamma-ray flashes (S-GRFs), are expected to be NS-NS mergers that do not lead to a BH, hence, without GeV emission. In \cite{2018arXiv180207552R}, it was shown that out of 18 S-GRFs, eight were not triggered by Fermi, and of the remaining ten sources, no GeV emission has been detected in the five sources observable by Fermi-LAT, i.e., with boresight angle $<75^\circ$. On the contrary, GeV emission has been observed in six S-GRBs: GRB 081024B, 090227B, 090510A, 140402A, 140619B, and 160829A (see Fig. \ref{fig:sGRBsFermiLAT}, reproduced from \cite{2018arXiv180207552R}). The rest-frame luminosity of these S-GRBs in the $0.1$--$100$ GeV energy band of Fermi-LAT shows a common power-law behavior of index $\approx -1.3$. Thus, we expect all S-GRBs to be explained by the NS-NS merger scenario and associated physical processes described here for GRB 090510.

For comparison, the GeV emission of BdHNe I has a power-law luminosity of index $\approx -1.2$ (see Fig. 3 in \cite{2021MNRAS.504.5301R}). It remains the relevant task for the future to determine whether this index difference in the GeV emission from the BH in S-GRBs and BdHNe I is physical, or if it will reduce with increasing S-GRB statistics. It is also worth noticing that the GeV emission of S-GRBs is systematically less luminous than that of BdHNe I \cite{2021MNRAS.504.5301R}. The GeV emission can, therefore, be a crucial discriminating factor to confirm the nature of sources with $T_{90}\approx 2$ s, i.e., around the traditional separatrix between short and long GRBs. This could be the case of GRB 240825A, which, with $T_{90} = 2.4$ s, could be the shortest ever observed BdHN I (Ruffini et al., submitted).

\appendix

\section{Dyadoregion of a rotating magnetic dipole}\label{app:A}

\citet{1973ApL....13..109R} obtained the solution of the Maxwell equations for a perfectly conducting sphere of radius $R$, endowed with a magnetic dipole of strength $B_p$ at the pole, rotating with angular velocity $\Omega$, and a uniformly distributed co-rotating charge, $Q$. This solution is a generalization of the well-known solution by \citet{1955AnAp...18....1D}. The electric and magnetic fields inside and outside the sphere are
\begin{subequations}\label{eq:EB}
    \begin{align}
    &(E_r,E_\theta,E_\phi) = \begin{cases}
        \displaystyle\left(-r \sin^2\theta \frac{\Omega}{c},-r \sin\theta\cos\theta \frac{\Omega}{c}, 0\right) \tilde B,& r<R\\
        \\
        \displaystyle\left(\frac{Q}{r^2} -P_2(\cos\theta)\frac{R^4}{r^4}\tilde\Omega \tilde B, -2\sin\theta\cos\theta \frac{R^4}{r^4} \tilde\Omega\tilde B ,0\right),& r\geq R
    \end{cases}\label{eq:E}\\
    &(B_r,B_\theta,B_\phi) = \begin{cases}
        \displaystyle\left(\cos\theta,-\sin\theta, 0\right) \tilde B,& r<R,\\
        \\
        \displaystyle\left(2\cos\theta,\sin\theta, 0\right) \frac{\tilde B}{2}\frac{R^3}{r^3},& r\geq R.
    \end{cases}\label{eq:B}
\end{align}
\end{subequations}
where $P_2 (\cos\theta) = (3\cos^2\theta-1)/2$ is the Legendre polynomial, and $\tilde B$ is the observed magnetic field 
\begin{equation}\label{eq:Btilde}
    \tilde B = B_p + \frac{2 \tilde\Omega}{3}\frac{Q}{R^2}.
\end{equation}
and we have defined the rotation parameter $\tilde \Omega \equiv \Omega R/c$.

The discontinuity of the electric field radial component at the surface leads to a surface charge density $\sigma$, which results in the total surface charge
\begin{equation}\label{eq:sigma}
    Q_{\rm tot} = \sigma 4 \pi R^2 = Q-\tilde \Omega B_p R^2 [P_2(\cos\theta)-\sin^2\theta] + {\cal O} (\tilde\Omega^2).
\end{equation}
Notice that it is given by the charge $Q$ and an \textit{effective charge} due to the Faraday-induced electric field, which on the pole ($\theta=0$) is $Q_{\rm eff} = -\tilde \Omega B_p R^2$.

The energy stored in the electromagnetic field is
\cite{1973ApL....13..109R}
\begin{align}\label{eq:W}
    W &= W_{\rm in} + W_{\rm ext} = \frac{1}{4}\int_0^\pi \int_0^R (E^2 + B^2) r^2 \sin\theta dr d\theta \nonumber \\
    &+\frac{1}{4}\int_0^\pi \int_R^\infty (E^2 + B^2) r^2 \sin\theta dr d\theta = \frac{Q^2}{2 R} +\frac{\tilde B^2 R^3}{4} \left(1 + \frac{2}{5}\tilde \Omega^2 \right)
\end{align}   
where we have used the axial symmetry of the field. In \cite{1973ApL....13..109R}, the value of the charge that minimizes the whole (interior+exterior) electromagnetic energy was calculated to be
\begin{equation}\label{eq:Qmin}
    Q_{\rm RT} = -\frac{1}{3} \tilde\Omega \tilde B R^2 = -\frac{1}{3} \tilde\Omega B_p R^2 + {\cal O} (\tilde\Omega^2),
\end{equation}
where RT refers as to the Ruffini-Treves value. Introducing the angular momentum, $J = I \Omega$, being $I$ the object's moment of inertia, the above charge acquires the perhaps more recognized form
\begin{equation}\label{eq:QwithI}
    Q_{\rm RT} = -\frac{1}{3\,k\, {\cal C}} \frac{c^3}{G} J B_p + {\cal O} (\tilde\Omega^2),
\end{equation}
where we have introduced the moment of inertia, $I = k M R^2$, being $k$ a form factor that accounts for the density distribution inside the object, and the compactness, ${\cal C} \equiv G M/(c^2 R)$. With this charge, the total surface charge (\ref{eq:sigma}) reads
\begin{equation}\label{eq:sigmamin}
    Q_{\rm tot}(\theta) = -\tilde \Omega B_p R^2 \left[ \frac{1}{3} + P_2(\cos\theta)-\sin^2\theta\right] + {\cal O} (\tilde\Omega^2),
\end{equation}
which, on the pole, is
\begin{equation}\label{eq:Qtot}
    Q_{\rm tot}(\theta=0) = 4 Q_{\rm RT} = -\frac{4}{3}\tilde\Omega B_p R^2 + {\cal O} (\tilde\Omega^2)= -\frac{4}{3\,k\, {\cal C}} \frac{c^3}{G} J B_p + {\cal O} (\tilde\Omega^2).
\end{equation}

We now calculate the observable magnetic field and energy for this charge value. By replacing the charge (\ref{eq:Qmin}) or (\ref{eq:QwithI}) into Eq. (\ref{eq:Btilde}), we obtain the observable magnetic field
\begin{equation}\label{eq:Btildemin}
    \tilde B = \frac{B_p}{1+\frac{2}{9}\tilde\Omega^2} = B_p \left( 1 - \frac{2}{9} \tilde\Omega^2\right) + {\cal O} (\tilde\Omega^2),
\end{equation}
while, from Eq. (\ref{eq:W}), we obtain the minimum electromagnetic energy
\begin{equation}\label{eq:Wmin}
   W_{\rm min} = \frac{B_p^2 R^3}{4} \left(1 + \frac{28}{45}\tilde \Omega^2 \right) + {\cal O} (\tilde\Omega^2).
\end{equation}

The comparison of Eq. (\ref{eq:Wmin}) and Eq. (\ref{eq:W}) in the case $Q=0$ tells us the charge contributes only at second order in the rotation parameter $\tilde\Omega$:
\begin{equation}\label{eq:Wdifference}
    \Delta \tilde W \equiv\frac{W_{\rm min} - W(Q=0)}{B_p^2 R^3/4} = \frac{2}{9}\tilde\Omega^2 + {\cal O} (\tilde\Omega^2).
\end{equation}

For the above parameters and $\Omega = 4054$ rad s$^{-1}$, $\tilde \Omega = 0.16$, so $\Delta \tilde W = 0.0058$. It is then clear that, even in the relatively fast rotating case we are treating, the linear approximation in $\tilde\Omega$ suffices quantitatively. Indeed, the above shows that, although conceptually relevant for the description of the system, the charge and the induced electric field contribute less than $1\%$ to the electromagnetic energy, which is dominated by the energy of the magnetic dipole, i.e., 
\begin{equation}\label{eq:Wmagf}
    W \approx W_{\rm mag} = \frac{1}{4} B_p^2 R^3.
\end{equation}

Having introduced the electromagnetic field of the configuration, we turn to define the region where spontaneous $e^+e^-$ pair creation can occur; the \textit{dyadoregion} \cite{2009PhRvD..79l4002C,2021PhRvD.104f3043M,2022EPJC...82..778R}). A coordinate-independent treatment of the pair creation in curved spacetime has been recently presented in Cherubini et al (2025, submitted) in the case of the Wald solution \cite{1974PhRvD..10.1680W}. We apply that framework to the present case of the rotating magnetic dipole with the electromagnetic field given in Eq. (\ref{eq:EB}), with $Q$ given by Eq. (\ref{eq:Qmin}).

First, we recall that the Schwinger pair production rate ($e^+e^-$ pairs per unit volume, per unit time) can be written as
\begin{equation}\label{eq:rate2}
	\dot{n} =\frac{e^2 \tilde E \tilde B}{4 \pi^2 \hbar^2 c}\sum_{n=1}^\infty\frac1{n}\coth\left(n\pi \frac{\tilde B}{\tilde E}\right)e^{-\frac{n\pi E_c}{\tilde E}},
\end{equation}
where ${\bf \tilde E}$ and ${\bf \tilde B}$ are the moduli of the electric and magnetic field in the frame where they are parallel. The latter can be calculated in terms of the electromagnetic invariants as $\tilde E=[({\mathcal F}^2+{\mathcal G}^2)^{1/2}-{\mathcal F}]^{1/2}$ and $ \tilde B=[({\mathcal F}^2+{\mathcal G}^2)^{1/2}+{\mathcal F}]^{1/2}$, being ${\cal F}\equiv\frac14F_{\mu\nu}F^{\mu\nu}$, $
{\cal G}\equiv\frac14F_{\mu\nu}{}^*F^{\mu\nu}$, with $F_{\mu \nu}$ and $^*F_{\mu \nu}$ the electromagnetic field tensor and its dual. From the exponential cutoff of the pair creation rate (\ref{eq:rate2}), it follows the natural definition of dyadoregion
\begin{equation}\label{eq:dyadoregion}
    \tilde{E}(r,\theta) = E_c,
\end{equation}
which implicitly sets the dyadoregion surface equation, $r = r_d(\theta)$. 

We now use that $\tilde E\approx E |\cos\theta|$ (Cherubini et al., 2025, submitted), where $E$ is the module of the electric field given by Eq. (\ref{eq:E}). Even with this simplification, Eq. (\ref{eq:dyadoregion}) is a quartic polynomial for $r_d$. The physical (real and positive) solution of the equation is
\begin{equation}\label{eq:rdya}
  r_d(\theta) = \sqrt{\frac{{\cal R} + {\cal D}}{2}},
\end{equation}
where ${\cal R} = \sqrt{z + 2a_2/3}$ and ${\cal D} = \sqrt{-{\cal R}^2 + 2 a_2 + 2 a_1/{\cal R}}$, with
\begin{equation}
    z = -2 \sqrt{\frac{p}{3}} \sinh{\left[\frac{1}{3}\text{arcsinh}\left(\frac{3 q}{2 p} \sqrt{\frac{3}{p}}  \right)\right]},
\end{equation}
being $p = (3 c - b^2)/3$, $q = (2 b^3-9 b c+27 d)/27$, $b = a_2$, $c=4 a_0$, $d = 4 a_0 a_2 - a_1^2$, having defined
\begin{align}
    a_0 &= \frac{1}{4}(\beta \tilde\Omega)^2 [(3 \cos^2\theta-1)^2 + \sin^2 2\theta] \cos^2\theta,\\
    a_1 &=\frac{1}{3}(\beta \tilde\Omega)^2 (3 \cos^2\theta-1) \cos^2\theta,\\
    a_2 &= \frac{1}{9}(\beta \tilde\Omega)^2 \cos^2\theta.
\end{align}
In the above equations, we have introduced the dimensionless magnetic field parameter $\beta\equiv B_p/B_c$. The dyadoregion width is $\Delta_d \equiv r_d(\theta)-R$. Thus, in general, to quantitatively define the dyadoregion size, we must set values for $B_p$ and $\tilde\Omega$ (so $\Omega$ and $R$).

\begin{figure}
    \centering
    \includegraphics[width=0.4\hsize,clip]{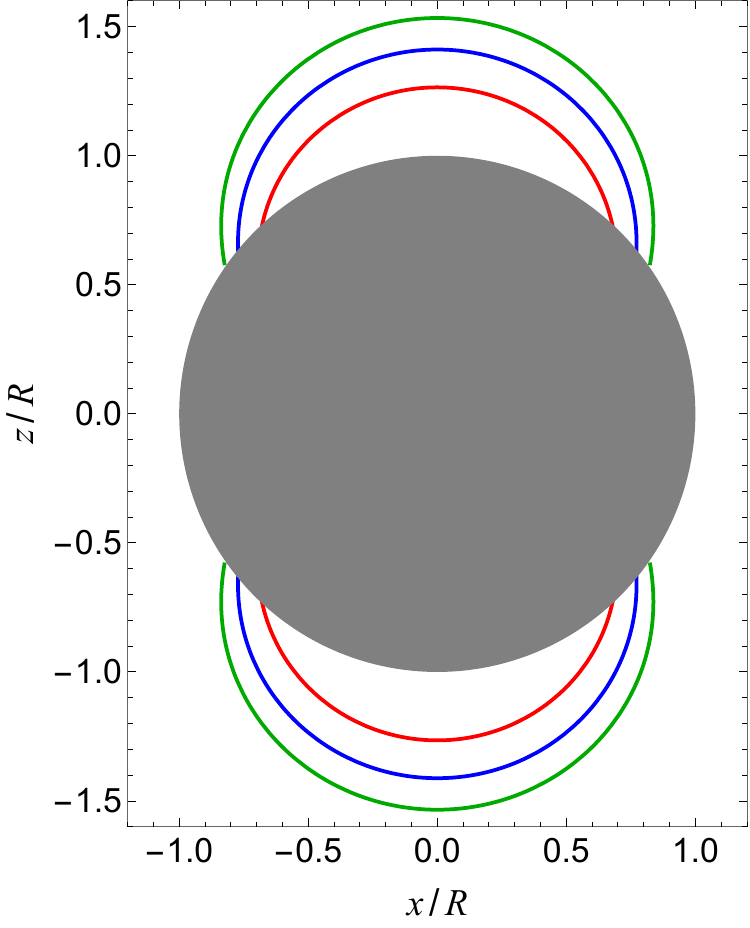}
    \caption{Dyadoregion given by Eq. (\ref{eq:rdya}), in the $x$-$z$ plane, for three values of the rotation parameter, $\tilde \Omega = 0.36$ (red), $0.51$ (blue), and $0.67$ (green). The magnetic field parameter is $\beta = 4.62$, i.e., $B_p = 2.04\times 10^{14}$ G. The coordinates are in units of the radius $R$. The merged object is the gray circle.}
    \label{fig:dyadoregion}
\end{figure}

Figure \ref{fig:dyadoregion} shows the dyadoregion of the Ruffini-Treves solution for three selected values of the rotation parameter, $\tilde \Omega = 0.36$ (red), $0.51$ (blue), $0.67$ (green), and magnetic field parameter $\beta = 4.62$, which corresponds to $B_p = 2.04\times 10^{14}$ G. It can be seen that the dyadoregion largest radial extension is along the polar axis. It is given by
\begin{equation}\label{eq:rdpole}
    r_d(\theta = 0) = R \sqrt{\frac{\beta \tilde\Omega}{6}} \left( 1 + \sqrt{1+\frac{36}{\beta \tilde\Omega}} \right)^{1/2}.
\end{equation}
Demanding that it be larger than $R$ (i.e., $\Delta_d > 0$) gives us the minimum magnetic field strength to produce pairs in the exterior
\begin{equation}\label{eq:Bmin}
    \beta_{\rm min} = \frac{B_{p,\rm min}}{B_c} = \frac{3}{4 \tilde\Omega}.
\end{equation}

The energy available for the pairs is the one stored in the dyadoregion:
\begin{equation}\label{eq:Epairs}
    {\cal E}_{e^+e^-} = \frac{1}{4} \int_0^\pi\int_R^{r_d(\theta)} (E^2 + B^2) r^2 \sin\theta dr d\theta.
\end{equation}
As the dyadoregion is formed by two polar lobes, its electromagnetic energy, which is the energy available for the pairs, is concentrated there. For the three examples shown in Fig. \ref{fig:dyadoregion}, the lobes cut the object surface ($r=R$) at $\theta\approx 43^\circ$ (red), $50^\circ$ (blue), and $55^\circ$ (green), respectively. Therefore, relative to an isotropic equivalent energy, the above result suggests a \textit{beaming factor} $f_b \equiv 1-\cos\theta_b \approx 1/3$.


\end{document}